# Facebook and political communication – Macedonian case

MSc. Sali Emruli[1], Mr.Tahir Zejneli[2] , MSc. Florin Agai[3]

**Faculty of Organization and Informatics, University of Zagreb,**
Varaždin, 42000, Croatia

**Faculty of Communication Sciences and Technologies, SEE-University,**
Tetovo, 1200, Macedonia

**Faculty of Electronics, Telecommunication and Informatics, University of Pristina,**
Pristina, Kosovo

## Abstract

Analysis how to use Internet influence to the process of political communication, marketing and the management of public relations, what kind of online communication methods are used by political parties, and to assess satisfaction, means of communication and the services they provide to their party's voters (people) and other interest groups and whether social networks can affect the political and economic changes in the state, and the political power of one party.

## 2. Introduction

The analysis will be done in a way that will create a list of largest political parties in the Republic of Macedonia, their communications infrastructure through ICT (website), content and the manner in which they placed their information and receive feedback from voters.

Internet, social networking, Web 2.0, Facebook, YouTube, blog ... All these are relatively new word in the political vocabulary, new concepts, new media and new opportunities for the transmission of ideas and messages are not enough channels used to communicate with the public. Although the practice of using the Internet in local political advertising goes back to the nineties, only in recent years the advent of new tools and social networks demonstrates true strength of this medium.

Besides direct access to the public, political ideas, it provides full force confrontation, but also provides a relatively convenient ground for review of public attitudes, research and development of certain ideas. Using such a change in social communication, transmission of political messages through the transition from traditional forms of communication and finding new paths to the recipients.

Professional and political public for years following the development of the Internet as a medium, but he showed the greatest strength in the last U.S. presidential election.

Political power depends on the satisfaction of the people towards a particular party and party connections with other parties or organizations. Well-developed social network provides further prestige and power of the party and its direct channel of communication with voters and other influential interests groups.

## 3. Internet and (political) communication in Macedonia

If we take as an example countries in transition and try to consider the possibilities of its integration into the global flow through the Internet, it is necessary to consider several questions:

What are the specific circumstances and how system works that is included in the new media developments?
Which are the main participants in the political and social events and the extent of openness to the network?
The extent to which effective new media elitism and the digital division, and how the public relates to the Internet?
Is there a digital ghettoization in society?

Macedonia is a small state in transition, but also the country that still bears the burden of the past. It is still more or less divided country, with many national, entities, local and regional government institutions and employees in them. Information and communication system is also not unique, and new media technologies are relatively slow to apply. We could say that Macedonia has entered into Internet galaxy in the "state of information confusion."

When it comes to pluralism (views, ideas, opinions), we can say that in Macedonia it is still in some way in the shadow of particularism. Specifically - still in force are divisions, and not the principle of multiple choices and when the information is in question.

The main carriers of political (and communication) processes are:

- International organizations that operate inside Macedonian system,
- Government institutions and
- Political parties, from which the cadres of government institutions are coming.

But, openness regard to the Internet is different as a new medium of communication and "relationship" to the public, in Macedonian, as well as worldwide community.

When Macedonia should insist on the integration into modern Europe, and inclusion in the contemporary global trends (political, as well as ICT), linking politics and the Internet should be a logical step in the development.

But the practice gives contrary arguments to the claim - internet is not fully involved in political communication in Macedonia (or not applied properly). Government institutions and political parties understand it as a bulletin board, so the network is mainly used to write pages that, basically, more encouraging monologue than dialogue. Such use of new technology aims to persuade rather than solve the problems, and websites are just digital versions of press releases, the party newsletters, etc. Most of the parties use the website to put on their advertising material (propaganda), after they have adopted to some extent the needs of the media. In other words, most parties use the web page for presentation (self advertising). The same is the case with most of public institutions. They mainly use the web site for presentations, announcements and messages. Content of their web page is maladjusted, or the pages are incomplete. Information mainly serves to encourage a positive image of institution. Such an attitude and a sense of self-sufficiency in the field of political communication on the Internet ignores the basic characteristic of the media, and it's ability to receive opinions and critical remarks of citizens, that is, getting an immediate response.

Macedonian political Internet space is the best proof of the thesis that politics at the website is structured experience and reflects the structure of organized political life in the real world with a relatively passive citizen. Oblak says Internet experienced "a shift from an unstructured, text-supported and interactive dialog practice to an organized monological web page, but in the Macedonian case it seems that the first three mentioned above had never been, or if they have any that was an exception, and not the rule. Although the interactivity on the essence of monological web sites exists; it seems that this interactivity is misunderstood. Specifically, is manifested in the so-called "testing the users' point of view ("Polling"), so many sites, particularly those of political parties have, "Question of the Day," "Customer Survey", etc. It often is incorrectly called the polling of public opinion, although methodologically unfounded, because respondents were so-called "self-selective" users (on questions, therefore, responses were only from those who have an interest in: visiting the web page, and then participate in the survey, and the sample is not representative)

However, such data is often manipulated and operated, so this is also cited in the traditional media. On the other hand, the electronic addresses of individuals, institutions and parties on the web sites usually are not updated, inaccurate or on the sent mails nobody responds.

When it comes to the Macedonian public, Internet only partially entered the wider use and mostly is dedicated to surf, chat, playing games and e-mail. This happens within the Macedonian society because the new media elitism and the digital division are in use, which is quite difficult to overcome because of the economic situation. Procurement of computer equipment and Internet connections are not expensive only to a small part of Macedonia's population that is employed and mostly live in cities.

Internet is not yet available everywhere and to everyone, although the steadily increasing number of ISPs (Internet Service Provider - a company that provides Internet access), the price still limits the expansion of the network. Those who have Internet access, it is mainly used one-sided. Users, rarely consider network as a global virtual place for interactive communication and exchange of views, in general users consider it as a "worlds hard disk drive" - Another in a series of projects for the outsourcing of memory means in which we find stored experiences, information and nothing more, as digital libraries, and most citizens if network is daily available to them, they only seek information on the Internet instead of publishing them. Only rare Macedonian users recognize the other very important "side of the coin" - interactivity and the ability to speak about their views. Is the fact that internet can contribute shaping public opinion on important issues of the Macedonian community and political realities, but there is a question of what the Macedonian emerging public Internet discusses. Even a quick look at some of the forum discussions, shows that they discuss mostly about less important issues and generally exchange messages that are totally with un-useful content. It's clear that, there are open endless possibilities for transferring the clear online public opinion in the offline world, but as long as the online public does not begin to deal with important political issues there's nothing to transfer in the offline public sphere.

## 4. Criteria for successful political advertising on social networks

There are three criteria that advertising campaigns on social networks need to fulfill in order to be successful: It has to be unobtrusive, creative and engage users.

**-Unobtrusive:** The inception of the internet saw the first iteration of online advertising in the form of pop-up ads: these opened new browsing windows flashing advertisements that would distract users from optimally experiencing the website.

However, consumer preferences saw increasing means to combat unwanted advertising. The advent of the pop-up blocker, which prevented any pop-up ads from opening on computers, forced advertisers to rethink their tactics. Thus, there has been a t rend towards online advertising becoming less obtrusive and more integrated with the look of the page. The advent of social networks has seen internet users becoming more and more empowered. They have free reign over how they portray themselves online, aided by the structure of the social networks themselves which allow them to capture their individuality online. They do not wish to be bothered by advertising at every possible opportunity, else they will leave. In fact, Facebook and Twitter are currently facing this problem, an that many users were dissatisfied with the high level of advertising on both sites and have stopped using them as a result. Thus, in order to be successful on social networks, future advertising campaigns have to be unobtrusive to ensure that users listen to their messages. Political parties are increasingly relying on the structure of social networks to spread the message among their users using technology savvy word of mouth techniques such as sharing articles, videos and applications.

**-Creative**: Advertising on social networks has to avoid traditional forms of online advertising, such as text-based ads and banner ads, in order to reach out to users. Social network website users are increasingly ignoring these forms of advertising as they reiterate similar messages and detract from usage experiences. In order to attract attention, companies need to deliver their message in imaginative ways that have never been done before. They should take advantage of the structure of social networks, such as applications on Facebook and Twitter, and the easy sharing between parties in order to spread awareness among users.

**-Engage users:** As mentioned earlier, social network websites have empowered users and allowed them to be creative. The technology behind the websites has enabled its users to fully display their unique personalities online. Companies can harness this expression of creativity by engaging the users in the advertising process themselves through social networks. This will give users a g reater sense of involvement with established political parties, eventually identifying themselves with the parties. Even if the political advertisements do not harness users " creativity, they should encourage user participation and involvement in order to develop a closer relationship.

## 5. Political Parties in Republic of Macedonia
## 5.1. Overview of the political system

Macedonia is a Republic having multi-party parliamentary democracy and a political system with strict division into legislative, executive and judicial branches. From 1945 Macedonia had been a sovereign Republic within Federal Yugoslavia and on September 8, 1991, following the referendum of its citizens, Macedonia was proclaimed a sovereign and independent state. The Constitution of the Republic of Macedonia was adopted on November 17, 1991, by the first multiparty parliament. The basic intention was to constitute Macedonia as a sovereign and independent, civil and democratic state and also to create an institutional framework for the development of parliamentary democracy, guaranteeing human rights, civil liberties and national equality.

The Assembly is the central and most important institution of state authority. According to the Constitution it is a representative body of the citizens and the legislative power of the Republic is vested in it. The Assembly is composed of 120 seats.

The President of the Republic of Macedonia represents the Republic, and is Commander-in-Chief of the Armed Forces of Macedonia. He is elected in general and direct elections, for a term of five years, and two terms at most.

Executive power of the Republic of Macedonia is bicephalous and is divided between the Government and the President of the Republic. The Government is elected by the Assembly of the Republic of Macedonia by a majority vote of the total number of Representatives, and is accountable for its work to the Assembly. The organization and work of the Government is defined by a law on the Government.

In accordance with its constitutional competencies, executive power is vested in the Government of the Republic of Macedonia. It is the highest institution of the state administration and has, among others, the following responsibilities: it p roposes laws, the budget of the Republic and other regulations passed by the Assembly, it determines the policies of execution of laws and other regulations of the Assembly and is responsible for their execution, decides on the recognition of states and governments, establishes diplomatic and consular relations with other states, proposes the Public Prosecutor, proposes the appointment of ambassadors and representatives of the Republic of Macedonia abroad and appoints chiefs of consular offices, and also performs other duties stipulated by the Constitution and law.

In Macedonia there are more political parties participating in the electoral process at national and local level.

### 5.2. Current Structure

Parties of traditional left and right:
Coalition VMRO – DPMNE (63 mandates, right oriented Macedonian party)
Democratic Party of the Albanians (12 mandates, right oriented Albanian party)
Coalition "SONCE" – SDSM (27 mandates left oriented Macedonian party)
Democratic Union for Integration (18 mandates left oriented Albanian party)

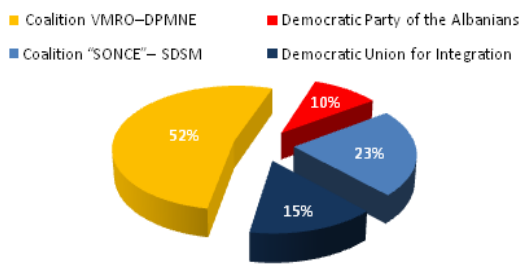

Fig. 1 Current political structure in Republic of Macedonia and Political Parties mandate percentage win in latest parliamentary elections.

### 6. Analysis: Political Parties on Social networks in Macedonia

The data on these parties' Web 2.0 presence was gathered through three types of sources. The first source was the official party web sites (Screenshots of political parties – Appendix), whose front pages were scanned for links to external Web 2.0 pages, as well as Web 2.0 elements built or embedded within the web site itself. This gives a good indication of what was officially approved by the party in question, in the sense of representing the party organization, as well as having a sufficient quality.

Most of the other parties were more cautious and advertised a s ingle element, either Facebook or blogs, although other data indicate they were present elsewhere as well.

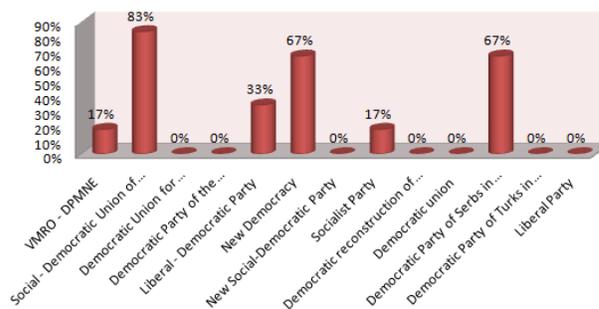

| Political Party | Facebook | Twitter | Blogs | Youtube | Flickr | Other | Total | Percentage |
|---|---|---|---|---|---|---|---|---|
| VMRO - DPMNE | 0 | 0 | 0 | 0 | 0 | 1 | 1 | 17% |
| Social - Democratic Union of Macedonia | 1 | 1 | 1 | 1 | 0 | 1 | 5 | 83% |
| Democratic Union for Integration | 0 | 0 | 0 | 0 | 0 | 0 | 0 | 0% |
| Democratic Party of the Albanians | 0 | 0 | 0 | 0 | 0 | 0 | 0 | 0% |
| Liberal - Democratic Party | 1 | 0 | 1 | 0 | 0 | 0 | 2 | 33% |
| New Democracy | 1 | 1 | 0 | 1 | 0 | 1 | 4 | 67% |
| New Social-Democratic Party | 0 | 0 | 0 | 0 | 0 | 0 | 0 | 0% |
| Socialist Party | 1 | 0 | 0 | 0 | 0 | 0 | 1 | 17% |
| Democratic reconstruction of Macedonia | 0 | 0 | 0 | 0 | 0 | 0 | 0 | 0% |
| Democratic union | 0 | 0 | 0 | 0 | 0 | 0 | 0 | 0% |
| Democratic Party of Serbs in Macedonia | 1 | 1 | 0 | 1 | 0 | 1 | 4 | 67% |
| Democratic Party of Turks in Macedonia | 0 | 0 | 0 | 0 | 0 | 0 | 0 | 0% |
| Liberal Party | 0 | 0 | 0 | 0 | 0 | 0 | 0 | 0 |

Table 1: Social networking elements on front page of parties' websites.

Fig.2 Social networking elements on political parties' websites front page

### 6.1. Facebook
### 6.1.1. Introduction

Facebook is a social networking site designed to connect users. Sites such as MySpace and Friendster are similar, but Facebook is generally considered the leading social networking site. Facebook allows individuals and organizations to create profiles that include personal interests, affiliations, pictures, and with some limitations virtually anything else a user or organization wants to post. Information entered in a profile links that user or organization to others who have posted similar information. For example, all political parties who list a particular ideology as a favorite or who share the same ideology constitutes a group. In user profiles, each of these pieces of data is a link; clicking on it d isplays everyone else in the network that included that element in their profiles. Every online political party needs to know how to use this social platform because facebook has grown to more than 400 million active users by May 2010. Its unique popularity in Macedonia is reflected in the fact that this tiny country with a population of 2.1 million has 708,960 active users, from which 408,220 (58.3%) are male and 291,660 (41.7%) female. According to age distribution for facebook means that 22% are minors (below 13 yrs – 17 yrs) without right to vote.

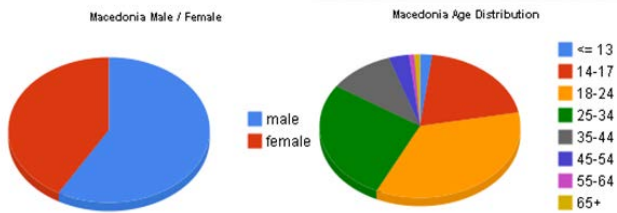

Fig. 3 Facebook statistics (gender and age)

Remain 78% (553,340) are political parties target audience which had the right to vote. In Macedonia right to vote have 1.792.082 citizens. From total number of Macedonian population 85% had the right to vote, from which 31% are active users of facebook.

Facebook's unprecedented growth apparently took the Macedonian party organizations by surprise, while its character as a social networking site for individual users gave them a particular challenge. However, individuals were allowed to set up special pages for particular groups and events, "to support or criticize another individual or entity". Parties could therefore establish party groups on Facebook, and they appeared to be protected by the passage in the user terms stating that "You may not set up a Facebook Page on behalf of another individual or entity unless you are authorized to do so". But they were to discover the implications of the statement that Facebook did not take any responsibility in this matter. Activists and sympathizers established groups to support their local party, local candidates, or even the national party. This led to a veritable jungle of groups, most of them using the party logo and linking to the party web site, but very few indicated whether they were the party's "official group" or not.

| | | |
|---|---|---|
| Total population | 2,100,000 | 100% |
| Population entitled to vote. | 1,792,082 | 85% |
| Total Facebook members | 708,960 | 40% |
| Facebook members entitled to vote. | 553,340 | 31% |
| Political party members on facebook | 157,762 | 29% |

Table 2: Number of users entitled to vote as members and fans of political parties groups and pages on Facebook.

Table 2: shows the available potential of voters on facebook and its absorption by political parties. From 31% facebook members entitled to vote only 29% (157 762) are participating in online political activities (members on party groups or fans of party pages).

Party informants knew Facebook was important for people belonging to the younger groups of voters, as well as activists, but the efficiency of Facebook regarding political networking they knew little or were skeptical about. Local and activist initiatives were appreciated, but there were little official activity on Facebook during the political campaigns.

In general Facebook was a party political anarchy, as there were big number of groups over which the national parties felt they had little oversight or control. Several parties discovered that a person already had established groups using the party name, the party logo etc. and henceforth appeared to an official group, but was not. In fact only two groups were established as "official", as the Centre Party group was first semi-official, and then sanctioned through a link on the party web site later. Furthermore, the party informants were aware of these groups and approved of their existence.

In order to determine whether there is a correlation between the behavior of voters and activities that can be monitored through social media, have conduct monitoring the activities of political parties through Facebook - most popular social network in Macedonia aimed for adults.

According to estimates portal IT.com.mk, in August 2008 Facebook had nearly 20,000 users from Macedonia, with a tendency for explosive growth which lead up to threefold increasing in just four months later. In December 2008 IT.com.mk estimate that Facebook has nearly 65 thousand users from Macedonia. In March 2009 IT.com.mk estimates that Facebook has about 150 thousand Macedonian users, which was confirmed next month with accurate assessment, according to which Facebook has 145,000 members from Macedonia or about 6 percent of the total population of the country.

Today if you compare the number of Facebook users in Macedonia we can conclude that it is increased by 4.9% for one year and today we have 708.960 users of Facebook or 29% of the total population of the country.

### 6.1.2. Limitations

Having regard to the protection of privacy as a basic value, especially in light of current fears in the Republic of Macedonia of various types of discrimination based on political determination, all observations were carried out only by groups, and not on the individual level. Therefore, were not made lists of which would be determined which person (user of Facebook) expressed support for certain political option.

For this and many other objective reasons arising from the nature of the medium, sample on which survey was conducted (persons who expressed support of any political parties) is self-selected, respectively is not structured in a way that reflects the demographic structure of the entire population of the RM.

As part of this approach should have in consideration the following restrictions:
Inability to determine whether each account of users on Facebook represents just one personality, and vice versa, whether certain people have open and manage more than one account.

Inability to determine whether all users of Facebook which took part in examination of means of communication are adult citizens of the Republic of Macedonia, respectively persons entitled to vote.

### 6.1.3. Methodology

As a main indicator for monitoring the activity on Facebook is used the number of users who are joined in one of the forms offered by political parties for self presentation and communication. Basically, were considered official profile, page and / or support groups. In comparison, as an indicator was taken into consideration the number of pages or number of groups and their supporters (number of group members and number of page fans).

### 6.1.4. Results

Table 21 shows the 13 Macedonian political parties which have available potential of voters on facebook and its absorption by political parties. From 31% facebook members entitled to vote only 29% (157 762) are participating in online political activities (members on party groups or fans of party pages). The following table represents the Facebook activity of 13 political parties in Republic of Macedonia. From where we can see that from 13 political parties that have been taken in this research all have groups or / and pages on Facebook.

Party that has the largest number of groups and pages on Facebook is the VMRO-DPMNE with a total of 137 groups or 67,991 members and 29 pages or 39,836 fans which confirms the credibility that has taken in the last parliamentary elections in 2008 where they have earned A total of 55 seats in Macedonia's parliament and this fact shows that VMRO-DPMNE is the largest political party in Republic of Macedonia. After the VMRO-DPMNE the SDSM is listed as second political party in the table with a total of 4 groups or 731 members and 23 pages or 29,098 fans, this party is also the second largest political party in Republic of Macedonia according the number of mandates won in the last parliamentary elections of 2008, which won 18 seats in Macedonia's parliament. The third party by the number of seats won in last parliamentary elections hold in 2008 is DUI with a total of 18 seats won on this parliamentary election in the Parliament of Republic of Macedonia, and according to its activity on Facebook is listed on six place with a total of 12 groups or 2601 members and 2 pages or 2132 fans.

DPA by the number of mandates won in the last parliamentary elections is ranked as fourth largest political party in Republic of Macedonia with six mandates that has won in recent elections of 2008 and for its activity on Facebook is listed on fifth place with a total of 11 groups or 2052 members and with 6 pages or 1967 fans.

Regarding the number of mandates won in the last parliamentary elections held in 2008 LDP is the fifth political party with a total of four seats in Macedonia's parliament and for its activity on Facebook is ranked in seventh place with a total of 3 groups or 695 members and 1 pages or two fans. After the LDP with the same number of four mandates in the Macedonians parliament is ND, which, by its activity on Facebook is ranked as fourth political party with a total of eight groups or 3309 members and 11 pages or 2620 fans. Regarding the number of earned seats on last Parliamentary elections of 2008 in Republic of Macedonia, NSD is ranked on six place based on the number of mandates won, and for its activity on Facebook NSD is listed in seventh place with a total of three groups or 1890 members.

Other parties like SP, DRM, DU, DPSM, DPTM and LP presented in the following table have the same political rating and sixth in the last parliamentary elections in Macedonia have won a seat in the parliament of Macedonia and by their activities on Facebook have the approximate number of groups and pages from which most distinguished is DPTM with 2 groups or 508 members and 3 pages or 153 fans, other political parties have just groups or just pages with an average of 400 members or fans.

| Political Party | Groups | Members | Pages | Fans |
|---|---|---|---|---|
| VMRO - DPMNE | 137 | 67991 | 29 | 39836 |
| Social - Democratic Union of Macedonia | 4 | 731 | 23 | 29098 |
| Democratic Union for Integration | 12 | 2601 | 2 | 2132 |
| Democratic Party of the Albanians | 11 | 2052 | 6 | 1967 |
| Liberal - Democratic Party | 3 | 695 | 1 | 2 |
| New Democracy | 8 | 3309 | 11 | 2620 |
| New Social-Democratic Party | 3 | 1890 | 0 | 0 |
| Socialist Party | 1 | 406 | 0 | 0 |
| Democratic reconstruction of Macedonia | 0 | 0 | 3 | 457 |
| Democratic union | 1 | 157 | 0 | 0 |
| Democratic Party of Serbs in Macedonia | 2 | 626 | 0 | 0 |
| Democratic Party of Turks in Macedonia | 2 | 508 | 3 | 153 |
| Liberal Party | 1 | 531 | 0 | 0 |
|  | 185 | 81497 | 78 | 76265 |

Table 3: Political parties' facebook groups their members, pages and their fans

| Political Party | % Groups | % Members | % Pages | % Fans | Average | Mandates |
|---|---|---|---|---|---|---|
| VMRO - DPMNE | 74% | 83% | 37% | 52% | 62% | 55 |
| Social - Democratic Union of Macedonia | 2% | 1% | 29% | 38% | 18% | 18 |
| Democratic Union for Integration | 6% | 3% | 3% | 3% | 4% | 18 |
| Democratic Party of the Albanians | 6% | 3% | 8% | 3% | 5% | 5 |
| Liberal - Democratic Party | 2% | 1% | 1% | 0% | 1% | 4 |
| New Democracy | 4% | 4% | 14% | 3% | 6% | 4 |
| New Social-Democratic Party | 2% | 2% | 0% | 0% | 1% | 3 |
| Socialist Party | 1% | 0% | 0% | 0% | 0% | 1 |
| Democratic reconstruction of Macedonia | 0% | 0% | 4% | 1% | 1% | 1 |
| Democratic union | 1% | 0% | 0% | 0% | 0% | 1 |
| Democratic Party of Serbs in Macedonia | 1% | 1% | 0% | 0% | 1% | 1 |
| Democratic Party of Turks in Macedonia | 1% | 1% | 4% | 0% | 2% | 1 |

Table 4: Political parties' facebook groups their members, pages, their fans and average of participation on facebook.

Table presents the activity of thirteen political parties on Facebook presented in percentage and averages of total activity and correlation with the number of seats acquired by every political party in recent parliamentary elections of 2008.

From the fig. below we can see the graphic presentation, which expresses activity of political parties on Facebook in percentage, groups and members of thirteen political parties obtained in the study.

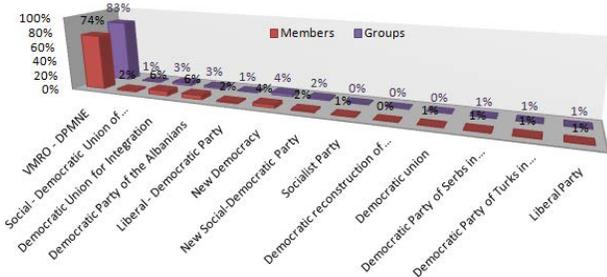

Fig.4 Political parties facebook groups and their members represented in percentage.

From the fig. below we can see the graphic presentation, which expresses activity of political parties on Facebook in percentage, pages and fans of thirteen political parties obtained in the study.

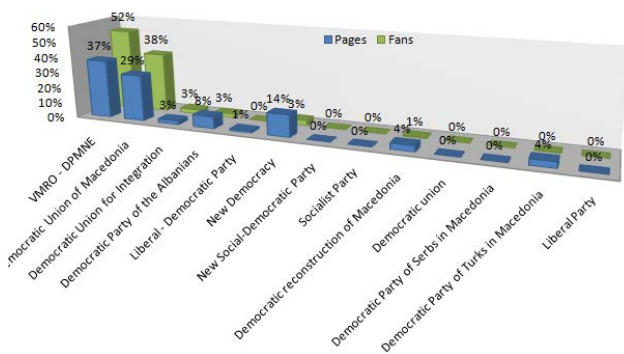

Fig.5 Political parties facebook pages and their fans represented in percentage

From the figure nr.6 we can see an average for every political party in correlation with the number of mandates won in the last parliamentary elections held in 2008.

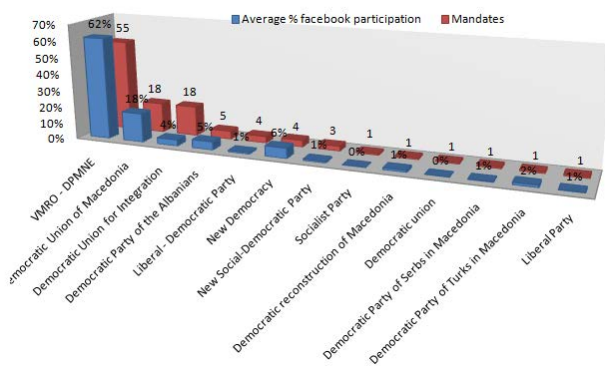

Fig.6 Political parties' participation average on facebook and their mandates.

## 7. Conclusion

From the results obtained from the research and the presentation of their graphics and table we can conclude that approximately all political parties with their Facebook activity give the same result compared to the number of mandates won in parliamentary elections. So VMRO-DPMNE leads in comparison with other parties, behind VMRO-DPMNE comes SDSM and DUI, after their coming DPA and ND and else. We can say that political parties have not yet exploit the potential offered from Facebook, observations can be made for the way how they use this social media, at the first point here is a question about the way of communication and self-advertising by political parties by using this social media, as they use it more as a means of information rather than as a media for interactive conversation with their electorate.

## 8. Reference

[1] Marin, Alexandra, and Barry Wellman. "Social Network Analysis: An Introduction." Computing in the Humanities and Social Sciences. Web. 04 Apr. 2010. <http://www.chass.utoronto.ca/~wellman/publications/newbies/newbies.pdf>.
[2] Oblak, Tanja. "Internet Kao Medij i Normalizacija Kibernetskog Prostora." Hrčak Portal Znanstvenih časopisa Republike Hrvatske. Web. 06 Apr. 2010. <hrcak.srce.hr/file/36810>.
[3] "Mixing Friends with Politics: A Functional Analysis of '08 Presidential Candidates Social Networking Profiles Authored by Compton, Jordan." All Academic Inc.

**S. Emruli,** received his bachelor degree from Faculty of Communication Sciences and Technologies in Tetovo SEE University (2006), MSc degree from Faculty of Organization and Informatics, Varaždin (2010). Currently works as professional IPA Advisor at Ministry of Local Self Government in Macedonia.

**T. Zejneli,** received his bachelor degree from Faculty of Communication Sciences and Technologies in Tetovo SEE University (2006). Currently works as Database administrator at Municipality of Tetovo in Macedonia.

**F. Agai,** received his bachelor degree from Faculty of Electronic Engineering in Skopje "St. Kiril and Metodij" University (1998), MSc degree from Faculty of Electronics, Telecommunication and Informatics in University of Pristina. Currently works as Professor at Electronics High School in Gostivar, Macedonia.